# Spin-orbit-splitting-driven nonlinear Hall effect in NbIrTe$_4$

Ji-Eun Lee[1,2,3,4], Aifeng Wang[5], Shuzhang Chen[5,6], Minseong Kwon[2,7], Jinwoong Hwang[1,8], Minhyun Cho[7], Ki-Hoon Son[2], Dong-Soo Han[2], Jun Woo Choi[2], Young Duck Kim[7], Sung-Kwan Mo[1], Cedomir Petrovic[5,6], Choongyu Hwang[3,*], Se Young Park[9,*], Chaun Jang[2,*], and Hyejin Ryu[2,*]

[1]*Advanced Light Source, Lawrence Berkeley National Laboratory, Berkeley, CA 94720, USA*

[2]*Center for Spintronics, Korea Institute of Science and Technology (KIST), Seoul 02792, Korea*

[3]*Department of Physics, Pusan National University, Busan 46241, Korea*

[4]*Max Planck POSTECH Center for Complex Phase Materials, Pohang University of Science and Technology, Pohang 37673, Korea*

[5]*Condensed Matter Physics and Materials Science Department, Brookhaven National Laboratory, Upton, New York 11973, United States*

[6]*Department of Physics and Astronomy, Stony Brook University, Stony Brook, New York 11794-3800, USA*

[7]*Department of Physics, Kyung Hee University, Seoul 02447, Korea*

[8] *Department of Physics, Kangwon National University,* Gangwondo *24341, Korea*

[9]*Department of Physics and Origin of Matter and Evolution of Galaxies (OMEG) Institute, Soongsil University, Seoul 06978, Korea*

*e-mail: ckhwang@pusan.ac.kr, sp2829@ssu.ac.kr, cujang@kist.re.kr, hryu@kist.re.kr

Current address: A. W.: School of Physics, Chongqing University, Chongqing 400044, China

## Abstract

The Berry curvature dipole (BCD) serves as a one of the fundamental contributors to emergence of the nonlinear Hall effect (NLHE). Despite intense interest due to its potential for new technologies reaching beyond the quantum efficiency limit, the interplay between BCD and NLHE has been barely understood yet in the absence of a systematic study on the electronic band structure. Here, we report NLHE realized in $NbIrTe_4$ that persists above room temperature coupled with a sign change in the Hall conductivity at 150 K. First-principles calculations combined with angle-resolved photoemission spectroscopy (ARPES) measurements show that BCD tuned by the partial occupancy of spin-orbit split bands via temperature is responsible for the temperature-dependent NLHE. Our findings highlight the correlation between BCD and the electronic band structure, providing a viable route to create and engineer the non-trivial Hall effect by tuning the geometric properties of quasiparticles in transition-metal chalcogen compounds.

Berry curvature (BC) is a key to understand novel physical phenomena such as anomalous Hall effect, chiral anomaly, topological Hall effect, and spin-valley Hall effect. Moreover, the BC classifies the topology of a solid via a topological number that predicts the presence of protected states at its boundary. The response of a system governed by the BC is constrained by the Onsager relation in the linear order[1]. This stringent constraint, however, is no longer valid in the high-order responses that are proportional to the second-order or even higher-orders of the driving field. The integration over higher-order fields gives a nonlinear response of the system, contributing to the optical and transport responses[2-5]. This finding not only provides a methodology to explore the momentum texture of the BC of a system, but also paves a way to utilize the response even though the linear order is vanishingly weak or when a large driving field makes the higher-order response exceed the linear order. To make use of the higher-order response has great potential for applications in rectification devices[6,7], photosensitive devices[8], and photovoltaic devices[9] that potentially overcome the quantum efficiency limit[10,11].

Recent studies on the higher-order response have invited nonmagnetic materials with broken centrosymmetry as a new member of the Hall effect family, so-called NLHE, characterized by a quadratic behavior of the Hall voltage with second-harmonic ($2\omega$) frequencies in the presence of a perpendicular AC driving current. In the absence of the linear Hall effect due to time-reversal symmetry, the lowest-order Hall current is driven by BCD[4]. Since the response is proportional to the gradient of the BC, tilted anticrossing bands and Weyl points[12-14] are predicted to exhibit strong BCD that can generate a nonlinear Hall angle close to 90 degrees[15]. As a result, the momentum-dependent texture of the BC based on the electronic structures is essential to understand the NLHE in which the evolution of BC momentum texture by lattice strain, interlayer twisting, and external electric fields[15] can lead to NLHE-based device applications[6,16-22]. Nevertheless, while the majority of studies on the NLHE have predominantly focused on exploring its transport properties, applications, and theoretical simulations, the direct experimental confirmation and comprehensive understanding of the BCD based on electronic structures have not yet been fully attained. This achievement would provide crucial insights into the underlying mechanism and controllability of the BCD.

In this paper, we report the room-temperature NLHE in NbIrTe$_4$ thin flakes which exhibit a frequency-doubled Hall conductivity proportional to the square of the driving current. We also demonstrate that the sign change in the NLHE at 150 K is induced by the sign change of the BCD because of the chemical potential shift at high temperatures. It is unambiguously evidenced by direct observation of a rigid shift in the temperature-dependent band dispersion using ARPES and calculated BCDs. Investigation on the electronic structures using ARPES and density functional theory (DFT) also indicates that the main contributor of BCDs is a partial occupation of spin-orbit split bands. Our findings provide important insights into the momentum texture of the Berry curvature and into controlling the Berry curvature dipole hosting the NLHE which can be utilized for NLHE-based devices.

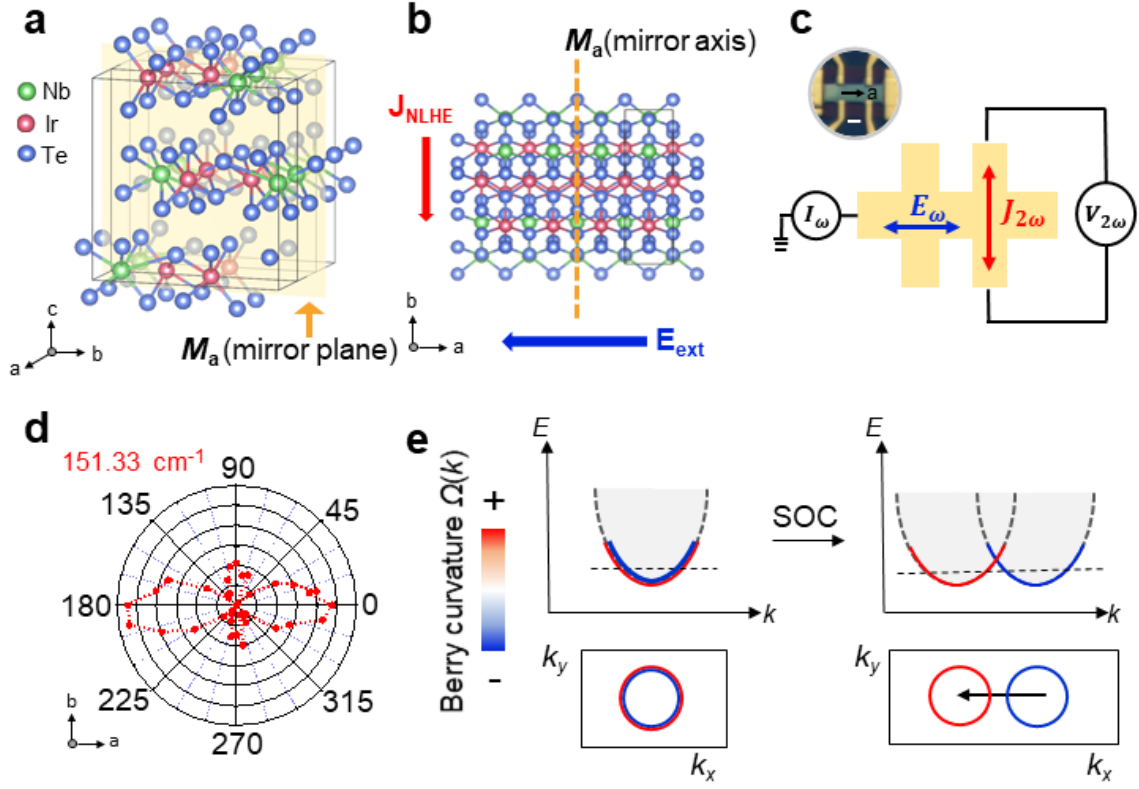

**Figure 1. Crystal structure symmetry and band model of NbIrTe$_4$ a**. Crystal structure of NbIrTe$_4$ exhibiting broken inversion symmetry with a mirror plane ($M_a$) illustrated as a yellow plane. Top view (**b**) of the crystal structure of NbIrTe$_4$ with mirror axis $M_a$ (yellow dashed line). $J_{NLHE}$ is the generated nonlinear Hall current parallel to $M_a$, and $E_{ext}$ is the applied electric field perpendicular to $M_a$. **c**. Illustrations of Hall devices with optical image (top left inset). $I_\omega$, $V_{2\omega}$, $E_\omega$, and $J_{2\omega}$ represent the current, voltage, electric field, and generated nonlinear Hall current, respectively, at frequencies $\omega$ and $2\omega$. The scale bar in the optical image (top left inset) is 1 μm. **d**. Polar plots of Raman modes at 151.33 cm$^{-1}$. The results are shown as intensity versus angle configuration. **e**. A schematic model picture of band structures with Berry curvature (BC) $\Omega(k)$ evolution including spin-orbit coupling (SOC). The lower panels display momentum-resolved BC, highlighting the emergence of a large dipole hotspot of BC in presence of SOC.

**Prediction of the nonlinear Hall effect in NbIrTe$_4$**

The crystal structure of bulk NbIrTe$_4$ (Fig. 1a) has an orthorhombic unit cell with space group $Pmn2_1$[23]. The experimental lattice constants are $a = 3.77$ Å, $b = 12.51$ Å, and $c = 13.12$ Å, where the two-dimensional planes are stacked along the $c$-axis. The material has a nonmagnetic metallic phase, and no anomalous Hall effect is expected. Moreover, because of the combination of mirror ($\{M_a|(0,0,0)\}$) and glide mirror ($\{M_b|(1/2,0,1/2)\}$) operations, no NLHE is expected within the *ab* plane[23-25]. However, in the slab geometry, the symmetry is lowered to *Pm* space group due to the breaking of the translation along the *c*-axis, leaving only identity and mirror ($\{M_a|(0,0,0)\}$) operations (Fig. 1b). The resultant symmetry allows nonlinear Hall current along the *b*-axis when there is a driving current along the *a*-axis, parallel to the direction of the BCD (Fig. 1b)[24,26]. The devices were fabricated into a Hall bars pattern (Fig. 1c), which is aligned in the crystallographic direction by angle-resolved polarized Raman spectroscopy (Fig. 1d). Crystallographic directions were confirmed via high intensity along the *a*-axis and low intensity along the *b*-axis, which is evidence of broken inversion symmetry in NbIrTe$_4$, consistent with the previous report[27]. In this system, the interplay between spin-orbit coupling and the presence of broken inversion symmetry leads to the emergence of a prominent BCD hot spot within spin-orbit-split bands (Fig. 1e), as elaborated upon in the subsequent theoretical section.

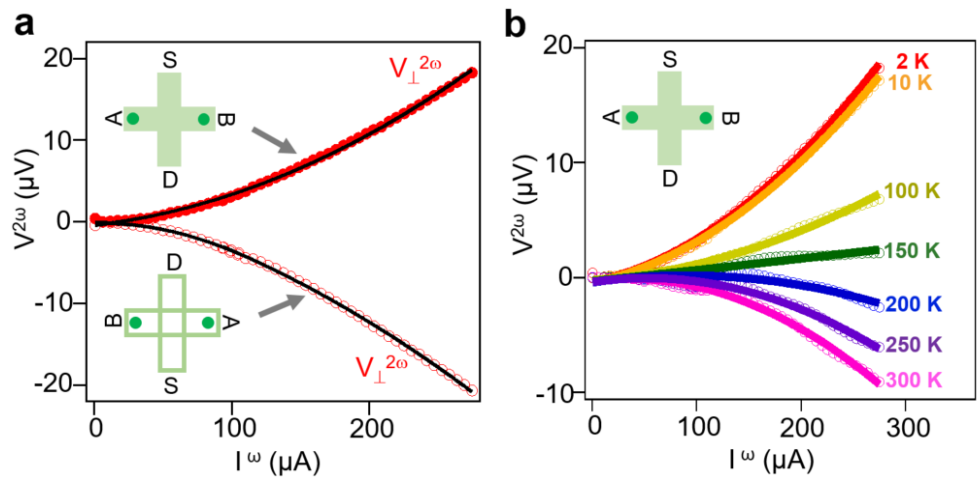

**Figure 2. Nonlinear Hall response in NbIrTe$_4$. a**. Second-harmonic $V_\perp^{2\omega}$ as a function of the AC current amplitude $I^\omega$ scaling quadratically in a 15 nm-thick flake of NbIrTe$_4$ at 2 K. The black solid line is a quadratic fit to the data. The current is along the *a*-axis. Green cross bars in the inset represent the geometry of the measurements. The applied current is injected from the source (S) electrode to the drain (D) electrode and the voltage is measured between A and B electrodes. A sign change occurs upon simultaneous reversal of both the applied current

direction and the corresponding Hall probes. **b**. Temperature dependence of the NLHE and a sizable NLH signal at room temperature, with a sign change at ~150 K.

**Observation of nonlinear Hall effect in NbIrTe$_4$**

To investigate the NLHE in NbIrTe$_4$, nonlinear transport measurements on a 15 nm-thick Hall bar device fabricated from a NbIrTe$_4$ flake were performed under zero magnetic field (Fig. 2). The second-harmonic transverse voltage ($V_\perp^{2\omega}$) under zero magnetic field in a 15 nm-thick NbIrTe$_4$ flake at 2 K responds quadratically to the current $I^\omega$ along the *a*-axis, indicating the presence of the NLHE on NbIrTe$_4$ flakes (Fig. 2**a**). Additionally, we have verified both the direction and frequency of driving AC current-dependent nonlinear Hall responses (see Fig. S2 and S3 of SI), thereby confirming the absence of experimental measurement artifacts. The second-harmonic transverse voltage ($V_\perp^{2\omega}$) of a NbIrTe$_4$ flake gradually decreases as the temperature increases from 2 to 150 K (nearly decay), followed by a sign change with an increasing magnitude as the temperature further increases to 300 K (Fig. 2**b**), consistent with the NLHE observed in TaIrTe$_4$[6]. However, the slight difference in the sign-changing temperature and the magnitude of NLHE between the two materials are due to the band structure changes associated with increased ionic radius and the spin-orbit coupling strength from Nb to Ta[12,15].

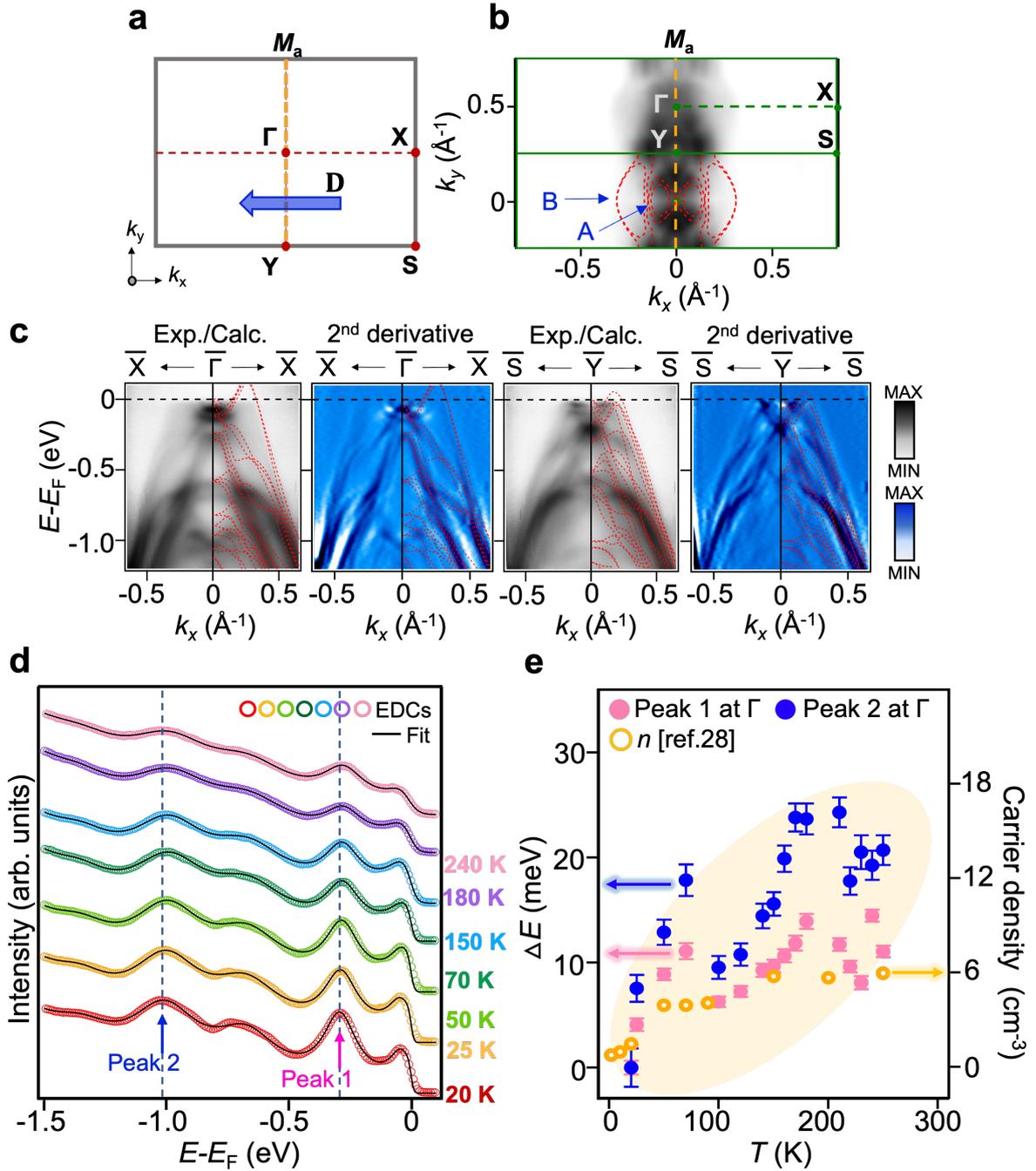

**Figure 3. Electronic structures and evidence for the temperature-dependent chemical potential shift of NbIrTe4. a.** The surface Brillouin zone of NbIrTe4 with high-symmetry points marked as red points. The blue arrow represents the BCD (D). **b.** ARPES and calculated (red dotted lines) Fermi surface (FS) of NbIrTe4 through two Brillouin zones (green solid lines) at 20 K. $M_a$ is a mirror plane. **c.** ARPES intensity plots with the calculated band structure (red dotted lines) and corresponding second-derivatives ARPES spectra for the enhanced visibility along the $\overline{X}$-$\overline{\Gamma}$-$\overline{X}$ and $\overline{S}$-$\overline{Y}$-$\overline{S}$ directions at 20 K. **d.** The energy distribution curves (EDCs) at the Γ point as a function of temperature along with black fitted curves. The multi-peak fits are

obtained by multiplication of Fermi-Dirac distribution function and convolution of instrumental resolution with Lorentzian curves (detailed on Fig. S7). Blue dotted lines are the position of the two EDC peaks at $E - E_F = -0.3, -1.0$ eV, respectively (where $E_F$ is the Fermi energy). **e.** Temperature dependence of the energy shifts $\Delta E$ (left axis) are taken from the peak position shift of EDCs at $\Gamma$ ($\Delta E = E - E_{T=20 \text{ K}}$) point obtained by multi-peak fit (see Fig. S7 in SI). Hole carrier density ($n$, right axis) is obtained from the reported Hall measurement[28].

**Temperature-dependent chemical potential shift nature of NbIrTe$_4$**

Since the NLHE is derived from BCD, understanding the nature of BCD from the perspective of electronic band structures enables both understanding and effective control of the NLHE. To investigate the electronic bands that dominantly contribute to NLH behavior, we explored the electronic band structure using ARPES experiments and DFT calculations. The Fermi surface of a NbIrTe$_4$ single crystal consists of elliptical-shaped electron pockets around the $\Gamma$ point (A in Fig. 3**b**) and semi-elliptical-shaped broad bulk hole pockets along the $\overline{\Gamma}$-$\overline{X}$ directions (B in Fig. 3**b**). Both Fermi surface features (Fig. 3**b**) and the band dispersions along the $\overline{X}$-$\overline{\Gamma}$-$\overline{X}$ and $\overline{S}$-$\overline{Y}$-$\overline{S}$ directions (Fig. 3**c**) are in good agreement with the DFT calculation results shown by red dotted lines in Fig. 3**b** and **c** and those reported previously[8,25,29-31].

To further elucidate the origin of the temperature dependence of the NLHE behavior (Fig. 2**b**), ARPES spectra along the $\overline{X}$-$\overline{\Gamma}$-$\overline{X}$ direction were acquired at various temperatures (Fig. S6 in SI). The energy distribution curves (EDCs) at the $\Gamma$ point as a function of temperature (Fig. 3**d**) demonstrate that, as the temperature increases, the two peaks at $E - E_F = -0.3, -1.0$ eV of the EDCs shifts systematically to a lower binding energy. In Fig. 3**e**, we show the temperature-dependent energy shift ($\Delta E$) for the peaks of EDCs at the $\Gamma$ point (blue, $\Delta E = E - E_{T=20 \text{ K}}$), obtained from multi-peak fits with multiplication of Fermi-Dirac distribution (FD) function and convolution of instrumental resolution as detailed in Fig. S7. It is clearly observed that the $\Delta E$ shifts about 15~20 meV as the temperature increases from 10 K to 280 K which implies the chemical potential shifts down to higher binding energy which is consistent with our simulated

ARPES spectra (See Fig. S11 in SI). The increase of the hole pockets in the Fermi surface associated with the chemical potential shift is consistent with the behavior of the hole carrier density, increasing as a function of temperature obtained from the reported transport results[28,30]. The ARPES and transport results suggest that there is substantial change in the band structures with increasing temperature. This could be related to the sign change in BCD which suggests the further investigation through DFT analysis.

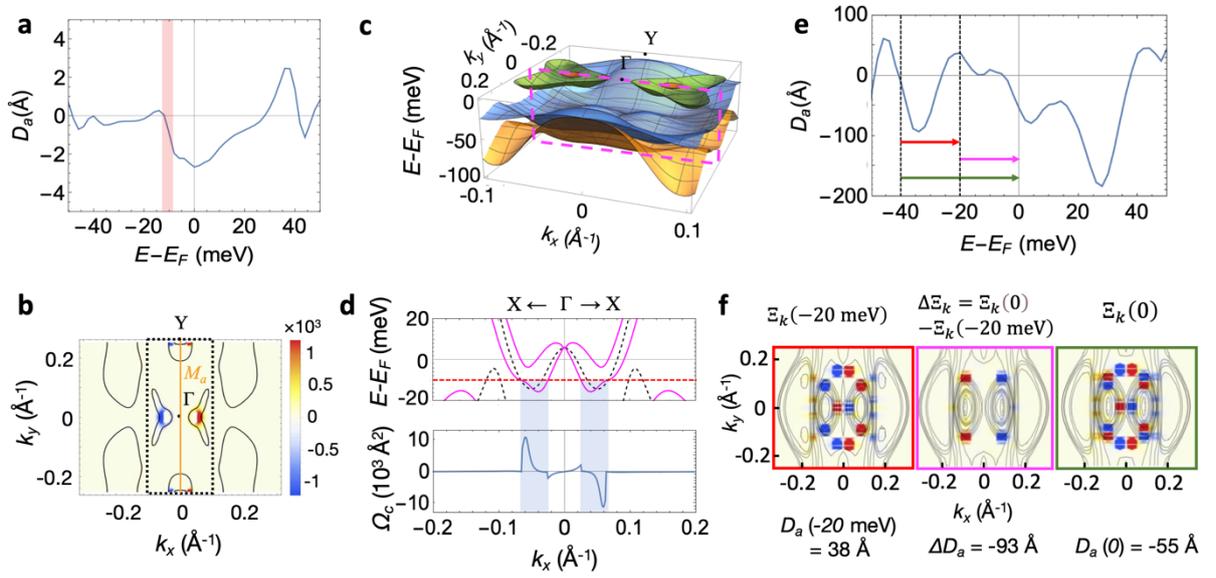

**Figure 4. Calculated Berry curvature dipole and momentum-dependent Berry curvature.**
**a.** The BCD ($D_a$) contributing to the in-plane NLHE for bilayer NbIrTe$_4$, plotted as a function of the chemical potential. **b.** Momentum-resolved BC ($\Omega_c(\mathbf{k})$ in Å$^2$) of bilayer NbIrTe$_4$ integrated from −12 to −7 meV corresponding to the red shaded area in panel **a**. **c.** Band structures in the area denoted as the black dotted square in panel **b**. **d.** Band dispersion around the $k$-points contributing large Berry curvatures of bilayer NbIrTe$_4$. The top panel shows bands near the Fermi energy expressed as solid magenta lines along the $\overline{X}$-$\overline{\Gamma}$-$\overline{X}$ cut (dotted magenta box in panel **c**. The black dashed lines are the bands calculated without spin-orbit coupling. The bottom panel is $\Omega_c(\mathbf{k})$ integrated up to −10 meV relative to $E_F$ (red dashed line). **e.** Chemical potential versus the component of the $D_a$ contributing to the in-plane NLHE for 12-layer NbIrTe$_4$. **f.** Integrated momentum-resolved BC ($\Xi_\mathbf{k}(E)$ for 12-layer NbIrTe$_4$ (arb. units) integrated from −40 meV to $E_F$, with the left, middle, and right panels corresponding to the energy range marked with red, magenta, and green arrows in panel **e**, respectively. The black lines denote iso-energy surfaces at $E_F$ − 20 meV (left), $E_F$ − 10 meV (middle), and $E_F$ (right).

**Origin of the BCD of NbIrTe$_4$**

We investigated the mechanism that induces the nonlinear Hall conductivity and the origin of the sign change by first-principles calculations. The current induced by the NLHE in two-dimensional systems is expressed as[4]

$$\boldsymbol{j}(2\omega) = \frac{e^3\tau}{2\hbar^2(1+i\omega\tau)}\hat{z} \times \boldsymbol{E}(\omega)\,(\boldsymbol{D}(\omega)\cdot\boldsymbol{E}(\omega)),$$

where $\hbar$ is Planck's constant, $\hat{z}$ is the direction normal to the plane parallel to the $c$-axis, $\tau$ is the scattering time, $\boldsymbol{E}(\omega)$ is the external electric field, and $\boldsymbol{D}(\omega)$ is the BCD vector. Given the symmetry of the NbIrTe$_4$ slab (space group $Pm$), $\boldsymbol{D}(\omega)$ is written as $D_a(\omega)\hat{a}$, where $\hat{a}$ is the unit vector along the $a$-axis, which gives the nonlinear current in the $b$-axis under electric fields applied along the $a$-axis. $D_a(\omega)$ is obtained by the derivative of the out-of-plane BC along the $a$-axis, expressed as

$$D_a(\omega) = \sum_n \int_{BZ} \frac{d^2k}{4\pi^2} f(\hbar\omega - \epsilon_{n\boldsymbol{k}}) \frac{\partial \Omega_{n\boldsymbol{k},c}}{\partial k_a},$$

where $n$ is the band index, $\epsilon_{n\boldsymbol{k}}$ denotes the single-particle energy, $f(\hbar\omega - \epsilon_{n\boldsymbol{k}})$ is the Fermi–Dirac distribution function, and $\Omega_{n\boldsymbol{k},c}$ is the Berry curvature along the $c$-axis. From the expression of BCD, the important parts of the Brillouin zone are regions with a large change in the BC along the $a$-axis. To identify the characteristics of the band structures that are the main contributors to the BCD, we first calculate $D_a(\omega)$ for a bilayer NbIrTe$_4$ slab. Compared with a thick slab geometry, which is difficult to analyze because of dense sub-bands, the bilayer system having a similar energy dependence of BCD with 12-layer NbIrTe$_4$ is suitable to pinpoint the hotspots of BC that mainly contribute to the BCD.

We find that the BCD is dominantly contributed from the partially occupied bands with spin-orbit induced splitting by analyzing the momentum texture of BC, integrated around the energy at which BCD changes abruptly. Fig. 4a shows the energy dependence of $D_a(\omega)$, where a large increase in $D_a$ is observed at around −10 meV. The momentum-resolved BC integrated in the energy range from −12 to −7 meV (red shaded area in Fig. 4**a**) shows that the BC is concentrated in small areas around the Γ point, with a clear sign change in $\Omega_{nk,c}$ across the $M_a$ mirror axis (Fig. 4**b**), which dominantly contributes to a sharp decrease in $D_a$. The bands contributing to the BC around −10 meV are presented in Fig. 4**c**, which corresponds to the region indicated by the black-dotted square in Fig. 4**b**. The one-dimensional cut along the $\overline{\text{X}}$-$\overline{\Gamma}$-$\overline{\text{X}}$ line reveals the characteristics of these bands (Fig. 4**d**), where large Rashba spin-orbit splitting is observed, consistent with the absence of inversion symmetry. The BC calculated along the $\overline{\text{X}}$-$\overline{\Gamma}$-$\overline{\text{X}}$ line (bottom panel of Fig. 4**d**) shows that partially occupied spin-split bands from spin-orbit coupling contribute largely to the BC ($\Omega_c$) (blue shaded area in Fig. 4**d**). Thus, we attribute the large change in the BCD to a shift of the $E_F$ as the occupation of the spin-orbit split bands changes accordingly. We note that the BCD and momentum-resolved BC of the bilayer slab are insensitive to the choice of the local density approximation (LDA) or generalized gradient approximation (GGA) for the exchange-correlation potential (see Fig. S12 in SI).

The sign change around the Fermi energy of the calculated BCD in a thicker slab (Fig. 4**e**) provides clear evidence that the sign inversion of the Hall conductivity can be induced by the Fermi energy shift. The calculated BCD for a 12-layer slab, which is sufficiently thick to represent the experimental geometry (see Fig. S8 in SI) shows a nonzero value of about $D_a$ = −60 Å at the Fermi energy. More importantly, the BCD changes sign as $E - E_F$ decreases, showing a positive peak about $D_a$ = 40 Å around −20 meV, which is comparable to the peak shift observed in the ARPES results (Fig. 3**d**). Thus, we propose that the observed sign change in the NLHE with increasing temperature is induced by a negative chemical potential shift, as evidenced by a sign change in the calculated BCD. Reducing the number of electrons corresponding to a 0.025 h/f.u. doping induces the chemical potential shifts about -20 meV both for bulk and slab geometry (see Fig. S12 in SI). We note that the $D_a$ from the experimental data is estimated to be −348 Å at low temperatures (see a section of 'Obtaining Berry curvature dipole from the transport data' in SI) which is similar to the theoretical value of −55 Å, as explained in the following discussion.

Figure 4**f** shows the change in the momentum texture of the BC of the 12-layer slab responsible for the energy-dependent sign change in BCD, where the dominant contribution to the change in the BC originates from small concentrated areas in the Brillouin zone, similar to the bilayer case. Since the BCD is nearly zero at −40 meV, we define the integrated BC in the energy interval from −40 meV to $E_F$ as $\Xi_k(E) = \sum_n \int_{E_F-40\,\text{meV}}^{E_F+E} dE' \, \delta(E' - \epsilon_{nk}) \, \Omega_{nk,c}$. Since the BCD along the $a$-axis $D_a(E)$ is proportional to $\frac{\partial \Omega_{nk,c}}{\partial k_a}$ summed over the bands below the energy $E$, large BCD is expected around the $k$-points showing an abrupt change in $\Xi_k(E)$ when integrated over a narrow energy range. Thus, $\Xi_k(-20\,\text{meV})$, $\Delta\Xi_k = \Xi_k(0) - \Xi_k(-20\,\text{meV})$, and $\Xi_k(0)$ (corresponding to red, magenta, and green arrows Fig. 4**e**, respectively) represent the momentum-dependent BC contributing $D_a$ at −20 meV (red box in Fig. 4**f**), $\Delta D_a$ between −20 meV and $E_F$ (magenta box), and $D_a$ at $E_F$ (green box), respectively. The positive BCD at −20 meV can be explained on the basis of an overall negative-to-positive sign change in the integrated BC across the mirror axis, giving a net positive value. The sign of BCD changes around $E_F$ as a result of the addition of the opposite component of the BCD shown in the middle panel with the opposite sign change in the BC across the mirror axis. Thus, the BCD at $E_F$ has a net negative sign with the BC texture, which is the sum of those corresponding to $D_a$ at −20 meV and $\Delta D_a$. From the analysis of the bilayer case, we expect that BCD of 12-layer slab can be also mainly contributed from the partially occupied spin-split bands. Our analysis reveals the mechanism of the temperature-dependent sign change in the

NLHE observed in the transport measurements. The shift in $E_F$, as observed in the temperature-dependent ARPES measurements, changes the distribution of the BC sensitive to band filling, resulting in the opposite sign of the BCD as the temperature increases.

**Conclusions**

We report the NLHE in non-magnetic NbIrTe$_4$ flakes. Our transport measurements show that the magnitude of the Hall voltage at a doubled frequency increases proportional to the square of the driving current, indicating the occurrence of the NLHE. Moreover, the measured Hall voltage shows a sign change with increasing temperature that persists up to room temperature. ARPES measurements and first-principles calculations reveal that the BCD in NbIrTe$_4$ originates mainly from partially filled spin-orbit split bands. Furthermore, we identified the mechanism of the sign changes in the NLHE on the basis of the chemical potential shift and associated change in the distribution of the BC in momentum space, which is supported by temperature-dependent energy shift in the ARPES results and sign-changing BC texture in the slab calculation. The identified mechanism suggests that the NLHE can be electronically controlled by changing the momentum-dependent texture of the BC, which can be used to enhance the efficiency of BCD-based devices such as memories and rectifiers.

**Methods**

**Device fabrication and electrical measurements.** NbIrTe$_4$ samples were mechanically exfoliated from a single crystal NbIrTe$_4$ grown as described elsewhere[32] and transferred onto SiO$_2$/Si substrates. Hall bar device geometries were patterned with Ti(10 nm)/Au(100 nm) electrodes on NbIrTe$_4$ crystals. To remove the oxidation layer on the surface of the NbIrTe$_4$ flakes, Ar$^+$-ion sputtering at 200 V was employed for 300 s. To perform nonlinear transport measurements, the device was biased with a harmonic current along the *a*-axis at a fixed frequency ($\omega$ = 117.77 Hz) and measured both the first-harmonic frequency ($\omega$) and second-harmonic frequency ($2\omega$) using a lock-in amplifier.

**Raman spectroscopy.** A linearly polarized 514 nm laser was focused to a spot of approximately 1-2 μm onto the nanoflakes at room temperature. The laser power was limited to less than 200 μW. A polar plot of angle-resolved measurements was obtained by fixing the Raman modes at 151.3 cm$^{-1}$ and comparing the intensity as a function of angle[33].

**ARPES measurements.** ARPES measurements were performed at Beamline 10.0.1, Advanced Light Source, Lawrence Berkeley National Laboratory. The ARPES system was equipped with a Scienta R4000 electron analyzer. The photon energy was set to be 60 eV, with an energy resolution of 20 meV and an angular resolution of 0.1 degrees.

**First-principles calculations.** We used the first-principles DFT to calculate the electronic structures and nonlinear Hall conductivity. The calculations were performed using the Vienna Ab-initio Simulation Package (VASP). The Ceperley-Alder (CA)[34,35] and Perdew–Burke–Ernzerhof (PBE)[36,37] parameterizations were used for the local density approximation (LDA) and the generalized gradient approximation (GGA), respectively. The projector augmented-wave method[38] was used with an energy cut-off of 500 eV. Spin-orbit coupling was included. For bulk and slab NbIrTe$_4$, *k*-point sampling grids of 3×12×3 and 3×12×1 were used, respectively. The experimental atomic structures were used to construct the bulk and slab geometry of 2- and 12-layer NbIrTe$_4$ with a vacuum layer of 16 Å. For the calculations of the BC and the BCD, the Wannier90 code[39] was used to construct the tight-binding Hamiltonian using Nb-*d*, Ir-*d*, and Te-*p* derived bands. For the bilayer, the tight-binding Hamiltonian was constructed from the bilayer-vacuum configuration. For the thicker slab, the tight-binding Hamiltonian was constructed by making a supercell of the bulk tight-binding Hamiltonian using the Python Tight Binding (PythTB) code[40]. The slab band structures of the supercell tight-binding Hamiltonian and those of DFT using vacuum-slab geometry were found to be in good agreement (Fig. S11 in SI). The calculations with a small hole doping of 0.025 h/f.u. were done by reducing the number of electrons with a compensating uniform background charge. The BC and BCD were calculated using the Wannier–Berri code[41,42] and the surface bands were calculated using WannierTools code[43].

## Supplementary Information

First harmonic Hall voltage of $NbIrTe_4$, The direction and frequency of driving AC current-dependent second harmonic Hall voltage of $NbIrTe_4$, Nonlinear Hall strength of $NbIrTe_4$ as a function of temperature, Determination of the thickness of the exfoliated $NbIrTe_4$ flake by atomic force microscopy (AFM), Temperature dependent electronic band structures of $NbIrTe_4$, Thickness dependence of the Berry curvature dipole, Obtaining Berry curvature dipole from the transport data, Comparison of slab band structures of supercell tight-binding and DFT calculations, Comparison of the band structures, BCD, and momentum-resolved BC between the GGA and LDA schemes, Determination of the surface termination and doping-dependent surface band structures.

## Acknowledgements

This work was supported by the KIST Institutional Program (2E32251, 2E31542), the National Research Foundation of Korea (NRF) grant funded by the Korea government (MSIT) (No. 2021R1A2C2014179, 2020R1A5A1016518, 2021R1A2C2011007, 2021M3H4A1A03054856, 2022M3H4A1A04096396, 2016K1A3A7A09005418, 2021R1A2C1004266, RS-2023-00284081, 2021R1C1C1009494, 2021R1A6A1A10044154, and 2022M3H4A1A04074153. This paper was also supported by the US DOE-BES, Division of Materials Science and

Engineering, under Contract No. DESC0012704 (BNL). The ARPES performed at the ALS was supported by the Office of Basic Energy Sciences, US DOE, under contract No. DE-AC02-05CH11231. J.-E. L. was supported by in part by an ALS collaborative Postdoctoral Fellowship. C. H. acknowledges support from the National Research Facilities and Equipment Center (NFEC) grant funded by the Ministry of Education (No. 2021R1A6C101A429). C. J. acknowledges support from Institute of Information & communications Technology Planning & Evaluation (IITP) grant funded by MSIT (No. 2022-0-01026). S. Y. P. acknowledges support from the Basic Science Research Program through the NRF funded by the Ministry of Education (No. 2021R1A6A1A03043957).

**Author contributions**

J.-E. L., C. H., S. Y. P., C. J., and H. R. proposed and designed the research. Y. L. and C. P. performed single-crystal growth. J.-E. L., M. K., and M. C. performed the transport and Raman spectroscopy measurements and analyzed the data with assistance from D.-S. H., Y. D. K., C. J., and H. R.; J.-E. L., J. H., and H. R. carried out the ARPES measurements and analyzed the ARPES data with assistance from C. H. and S. K. M.; S. Y. P. carried out the density functional calculations and provided theoretical support. J. -E. L., S. Y. P., C. J., and H. R. wrote the manuscript and revised it with assistance from K. S., C. H., J. W. C., C. P., and S. K. M. All authors contributed to the scientific planning and discussions.

**Additional information**

SI is available in the online version of the paper. Reprints and permissions information is available online at www.nature.com/reprints. Correspondence and requests for materials should be addressed to C. H. (ckhwang@pusan.ac.kr), S. Y. P. (sp2829@ssu.ac.kr), C. J. (cujang@kist.re.kr), or H. R. (hryu@kist.re.kr).

**Competing financial interests**

The authors declare no competing financial interests.

Supplementary Information for

Spin-orbit-splitting-driven nonlinear Hall effect in NbIrTe$_4$

Ji-Eun Lee[1,2,3,4], Aifeng Wang[5], Shuzhang Chen[5,6], Minseong Kwon[2,7], Jinwoong Hwang[1,8], Minhyun Cho[7], Ki-Hoon Son[2], Dong-Soo Han[2], Jun Woo Choi[2], Young Duck Kim[7], Sung-Kwan Mo[1], Cedomir Petrovic[5,6], Choongyu Hwang[3,*], Se Young Park[9,*], Chaun Jang[2,*], and Hyejin Ryu[2,*]

[1]*Advanced Light Source, Lawrence Berkeley National Laboratory, Berkeley, CA 94720, USA*

[2]*Center for Spintronics, Korea Institute of Science and Technology (KIST), Seoul 02792, Korea*

[3]*Department of Physics, Pusan National University, Busan 46241, Korea*

[4]*Max Planck POSTECH Center for Complex Phase Materials, Pohang University of Science and Technology, Pohang 37673, Korea*

[5]*Condensed Matter Physics and Materials Science Department, Brookhaven National Laboratory, Upton, New York 11973, United States*

[6]*Department of Physics and Astronomy, Stony Brook University, Stony Brook, New York 11794-3800, USA*

[7]*Department of Physics, Kyung Hee University, Seoul 02447, Korea*

[8] *Department of Physics, Kangwon National University,* Gangwondo *24341, Korea*

[9]*Department of Physics and Origin of Matter and Evolution of Galaxies (OMEG) Institute,*

*Soongsil University, Seoul 06978, Korea*

*\*e-mail: ckhwang@pusan.ac.kr, sp2829@ssu.ac.kr, cujang@kist.re.kr, hryu@kist.re.kr*

*Current address: A. W.: School of Physics, Chongqing University, Chongqing 400044, China*

**First harmonic Hall voltage of NbIrTe$_4$**

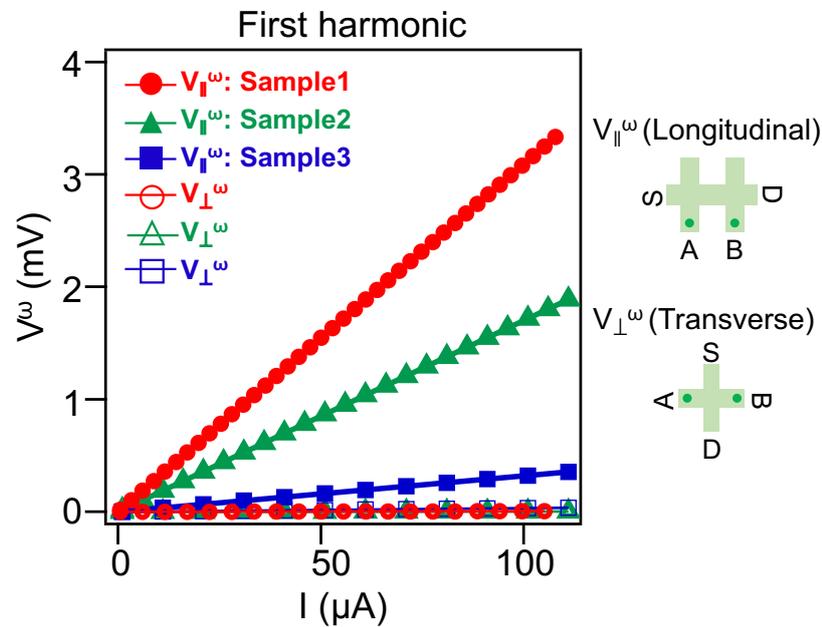

**Figure S1**. First-harmonic $V_\parallel^\omega$ (longitudinal) and $V_\perp^\omega$ (transverse) as function of the AC current $I^\omega$ for three different samples at 2 K. (Sample 1, Sample 2, and Sample 3). All measured $V_\parallel^\omega$ values (closed symbols) consistently exceed the corresponding $V_\perp^\omega$ values (open symbols), indicating the disappearance of the linear Hall response ($V_\perp^\omega$) due to preserved time-reversal symmetry. The green crossbars illustrate the experimental geometries. $V_\parallel^\omega$ and $V_\perp^\omega$ represent the first-harmonic voltage parallel and perpendicular to the applied current direction (four probe *IV* measurement and Hall measurement), respectively. The applied current is injected from the source (S) electrode to the drain (D) and the voltage is measured between A and B electrodes.

First harmonic voltages of NbIrTe$_4$ are measured for three different flakes at 2 K. The longitudinal voltage $V_\parallel^\omega$ responds linearly with the current $I^\omega$ along the *a*-axis, identical to the behavior observed in the four-probe *I–V* measurement result, whereas the transverse voltage $V_\perp^\omega$ (Hall voltage) is much smaller than $V_\parallel^\omega$ across all three samples. Due to the preserved time-reversal symmetry, the linear Hall response ($V_\perp^\omega$) should vanish due to the zero BC nature[1].

**The direction and frequency of driving AC current-dependent second harmonic Hall voltage of NbIrTe$_4$**

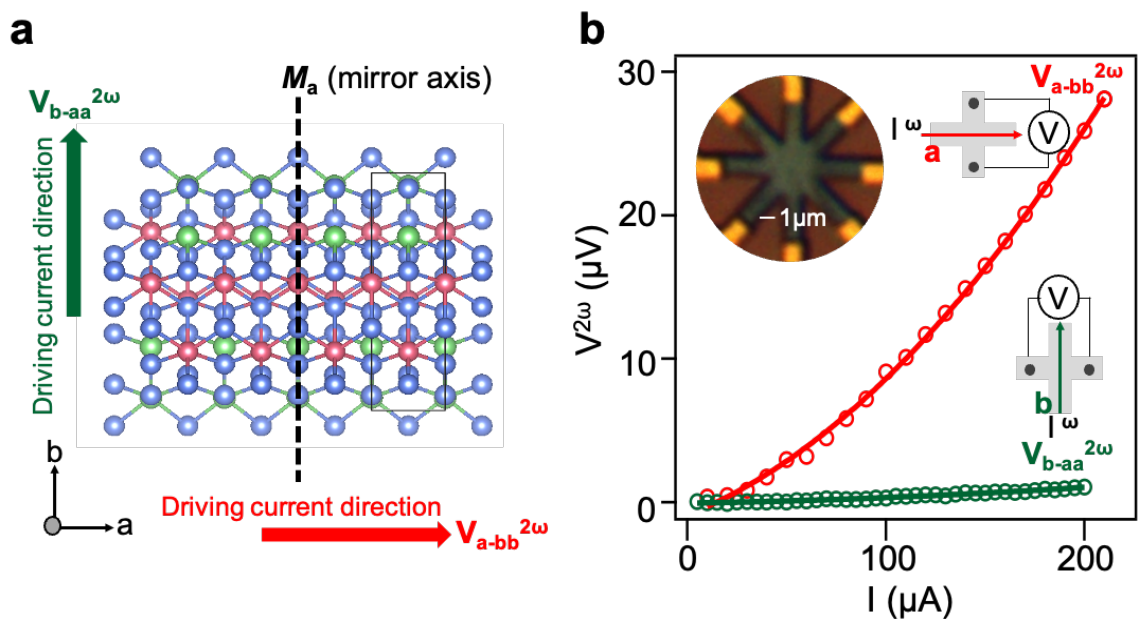

**Figure S2**. The direction of driving AC current direction-dependent nonlinear Hall voltage ($V^{2\omega}$) in NbIrTe$_4$. (a) The in-plane crystal structure with unit cell of square box. The mirror axis is parallel to *b*-axis and perpendicular to *a*-axis. The AC current is applied along *a*- and *b*-axis, labeled as $V_{a\text{-}bb}^{2\omega}$ and $V_{b\text{-}aa}^{2\omega}$. (b) The second-harmonic transverse voltage depending on driving

current direction along the different two crystal axes. The inset shows an optical microscopic image of the measured device references with 1 μm scale bars for reference.

We measured second harmonic transverse Hall voltage as varying current direction along the two different crystal axes i.e., a-axis and b-axis in Fig. S2. Considering crystal symmetry, only the driving current along a-axis should exhibit the NLHE signal. Evidently, the second-harmonic Hall voltage ($V_{a\text{-}bb}^{2\omega}$) with driving current along the a-axis exceedingly surpass the second-harmonic Hall voltage ($V_{b\text{-}aa}^{2\omega}$) with driving current along the b-axis, suggesting that there is the absence of NLHE along b-axis.

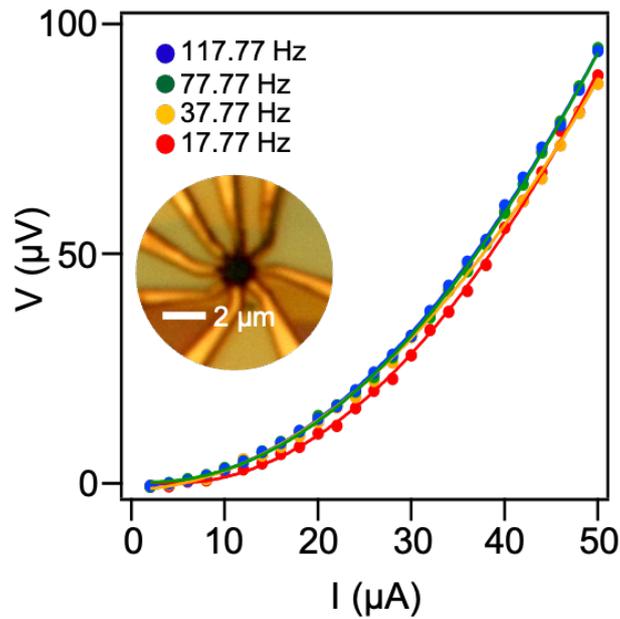

**Figure S3.** Nonlinear Hall voltage ($V^{2\omega}$) for several different frequencies of driving AC current in NbIrTe$_4$. The second-harmonic voltage for different driving current frequencies of 117.77 Hz, 77.77 Hz, 37.77 Hz, and 17.77 Hz. The inset shows an optical microscopic image of the measured device with a 1 μm scale bar.

We also measured second harmonic transverse Hall voltage by changing frequencies in Fig. S3. We clearly observe no frequency dependence in the measurement, suggesting that we can exclude measurement artifacts.

**Nonlinear Hall strength of NbIrTe$_4$ as a function of temperature**

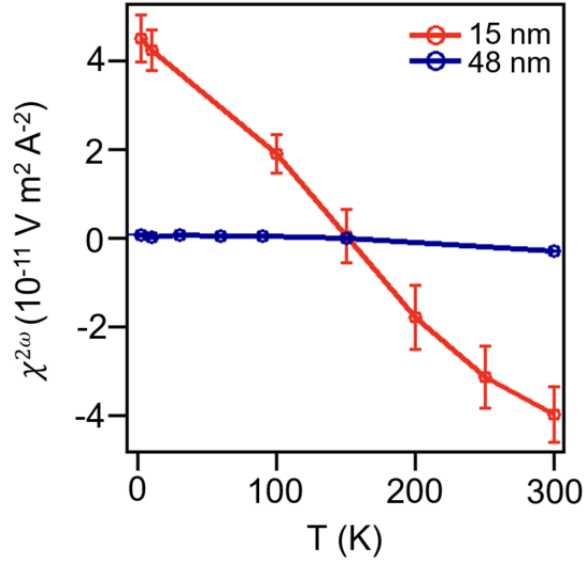

**Figure S4.** The strength of NLHE at different thickness of flakes, 15 nm and 48 nm, as a function of temperature.

We performed NLHE measurements on NbIrTe$_4$ flakes with two different thicknesses, 15 nm and 48 nm, to compare the NLHE in different thickness. The strength of the nonlinear Hall conductivity, $\chi^{2\omega}$, indicates a quadratic coefficient of the parabolic fitting function of $V_\perp^{2\omega}$ vs. $I^\omega$. The sign of the NLH signal changes at ~150 K irrespective of the sample thickness, whereas the NLH signal $\chi^{2\omega}$ for 15 nm-thick flakes is substantially higher than that for 48 nm-thick flakes, supporting the previous argument that the NLH signal can only originate from the surface because of the absence of the glide mirror plane on the surface[2].

**Determination of the thickness of the exfoliated NbIrTe$_4$ flake by atomic force microscopy (AFM)**

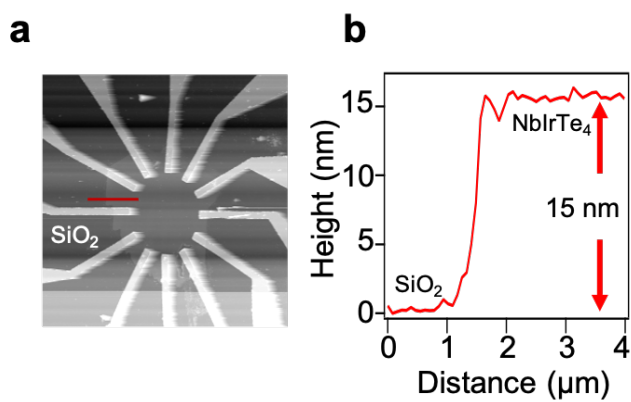

**Figure S5.** (a) An AFM image showing a 15 nm-thick flake of NbIrTe$_4$ on an SiO$_2$ substrate. (b) The measured height profile along the red line in panel a.

We confirmed the thickness of an exfoliated NbIrTe$_4$ using atomic force microscopy (AFM). The AFM image is shown in Fig. S5 (a), revealing the 15 nm-thick of NbIrTe$_4$ flake on SiO$_2$ substrate. The height profile of the NbIrTe$_4$ flake in Fig. S5 (b) was taken along the red line in (a).

**Temperature-dependent electronic band structures of NbIrTe$_4$**

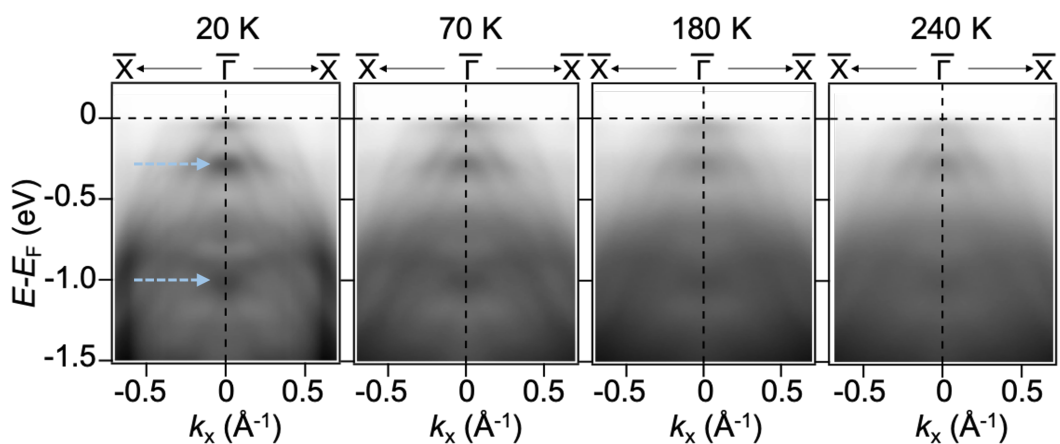

**Figure S6.** Temperature-dependent ARPES intensity cuts along the $\overline{X}$-$\overline{\Gamma}$-$\overline{X}$ direction at temperature of 20, 70, 180, 240 K, respectively. Three horizontal arrows indicate the position of the three EDC peaks at $E$-$E_F$ = -0.3, -1.0 eV, respectively.

We show representative temperature-dependent ARPES *E-k* cuts along Γ-X direction at temperature of 20 K, 70 K, 180 K, and 240 K, respectively. In Fig. 3(d), EDCs are extracted from the black dashed line at Γ point in Fig. S6.

**Multi-peak fit analysis of temperature-dependent EDCs**

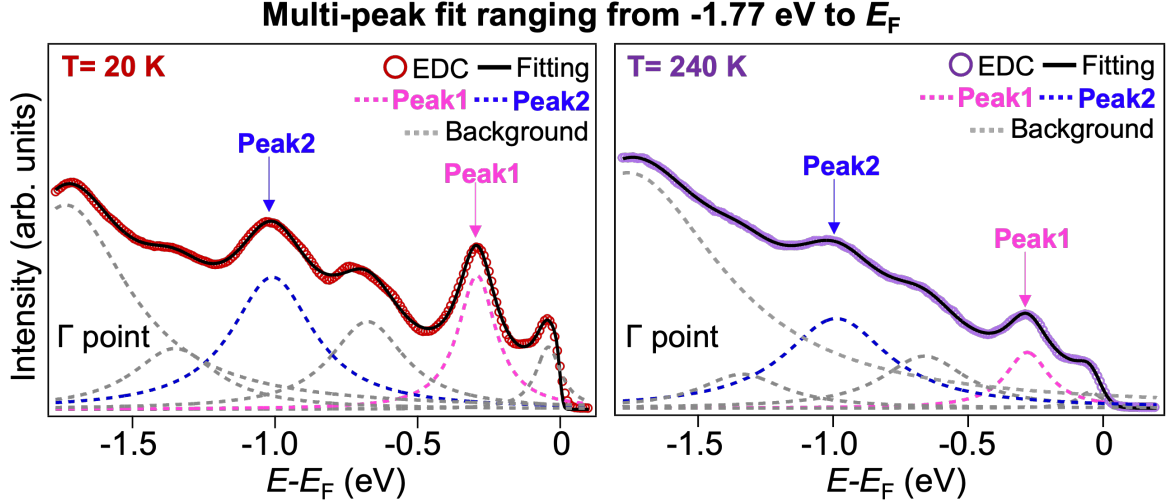

**Figure S7**. Multi-peak fits of energy distribution curves (EDCs) at the Γ, which is fitted with Lorentzian peaks, incorporating the multiplication of the Fermi-Dirac distribution (FD) function and convolution with instrumental resolution. The energy range of fitting is from -1.77 eV to Fermi energy ($E_F$) with temperatures ranging from 20 K to 240 K.

Our fitting analysis involves the Lorentzian peaks to cover a wide range of energy in the Γ point, which is achieved by the multiplying the FD function and convoluting the instrumental resolution. Now, the fitted curves are accurately aligned with the Fermi level cut-off energy, as presented in Figure S7. Furthermore, when the temperature increases to 240 K (Γ point), distinct peaks labeled as Peaks 1 and 2 (Γ point) are clearly resolved. We note that other peaks denoted with grey dotted lines in Figure S7 exhibit significant broadening at higher temperatures, displaying fluctuating weight during the fitting process, rendering them unreliable.

**Thickness dependence of the Berry curvature dipole**

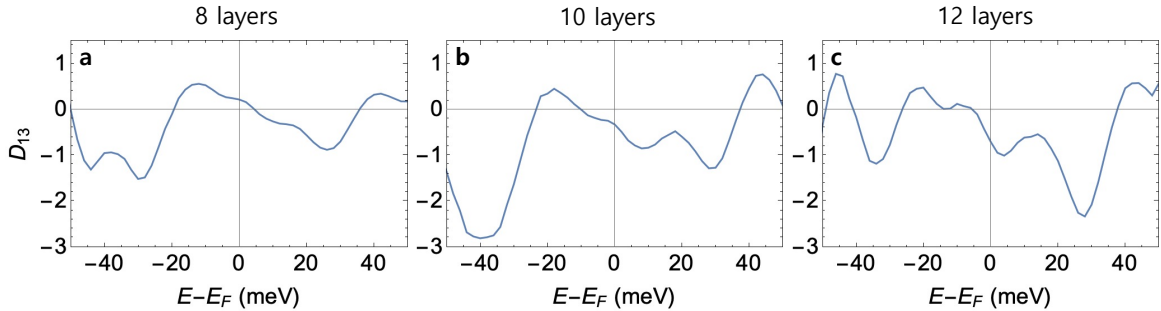

**Figure S8.** The calculated Berry curvature dipoles for supercells with various thicknesses of 8, 10, and 12 layers.

As the Berry curvature dipole (BCD) is calculated in a slab geometry with a thickness smaller than the thickness of an experimental sample, we checked the dependence of the calculated BCD as a function of slab thickness. Fig. S8 shows the BCD calculated for slabs thicknesses of 8, 10, and 12 layers; further increasing the number of slabs is challenging because of the large numbers of Wannier functions required for calculating the BCD (1056 Wannier functions for 12 layers). We find that the key feature in the BCD (i.e., a positive-to-negative sign change as the chemical potential is increased) is maintained with increasing thickness. Thus, we expect that the same feature will be maintained for the thicker slab.

**Obtaining Berry curvature dipole from the transport data**

We estimate the BCD using the following formula from Eq (1):

$$j_b(2\omega) = \frac{e^3 \tau}{2\hbar^2(1+i\omega\tau)} D_a(\omega)|E_a(\omega)|^2,$$

where $\hbar$ is Planck's constant, $\tau$ is the scattering time, $E_a(\omega)$ is an external electric field along the $a$-axis, $D_a(\omega)$ is the BCD along the $a$-axis, and $j_b(2\omega)$ is the second-harmonic current density along the $b$-axis. We assume the static limit ($\omega\tau \ll 1$)[3] and express $D_a(\omega)$ as

$$D_a(\omega) = \frac{2\hbar^2 j_b(2\omega)}{e^3 \tau |E_a(\omega)|^2} = \frac{2\hbar^2 n t^2 W V_b(2\omega)}{e\, m^* \rho_0^2\, |I_a(\omega)|^2}$$

where $n$ is the charge density, $t$ and $W$ are the thickness and width of the sample, respectively, $V_b(2\omega)$ is the second-harmonic Hall voltage along the $b$-axis, $m^*$ is the effective mass that we approximate as the electron mass ($m_e$), $I_a(\omega)$ is the current along the $a$-axis, and $\rho_0$ is the resistivity in the static limit, given as $\frac{m^*}{n e^2 \tau}$. By substituting $n = 2.1 \times 10^{20}$ cm$^{-2}$, $t = 30 \times 10^{-7}$ cm, $W = 5 \times 10^{-7}$ cm, $V_b(2\omega) = 10$ μV, $\rho_0 = 30$ μΩ/cm², and $I_a(\omega) = 60$ μA, we obtain $D_a(\omega) = 4.43$. Since the BCD in the two-dimensional limit is compared between theory and experiments, the relevant value is $D_a(\omega) \times t_{\text{calc}} = 348$ Å, where $t_{\text{calc}}$ is the thickness of the 12-layer supercell (78.5 Å).

**Comparison of slab band structures of supercell tight-binding and DFT calculations**

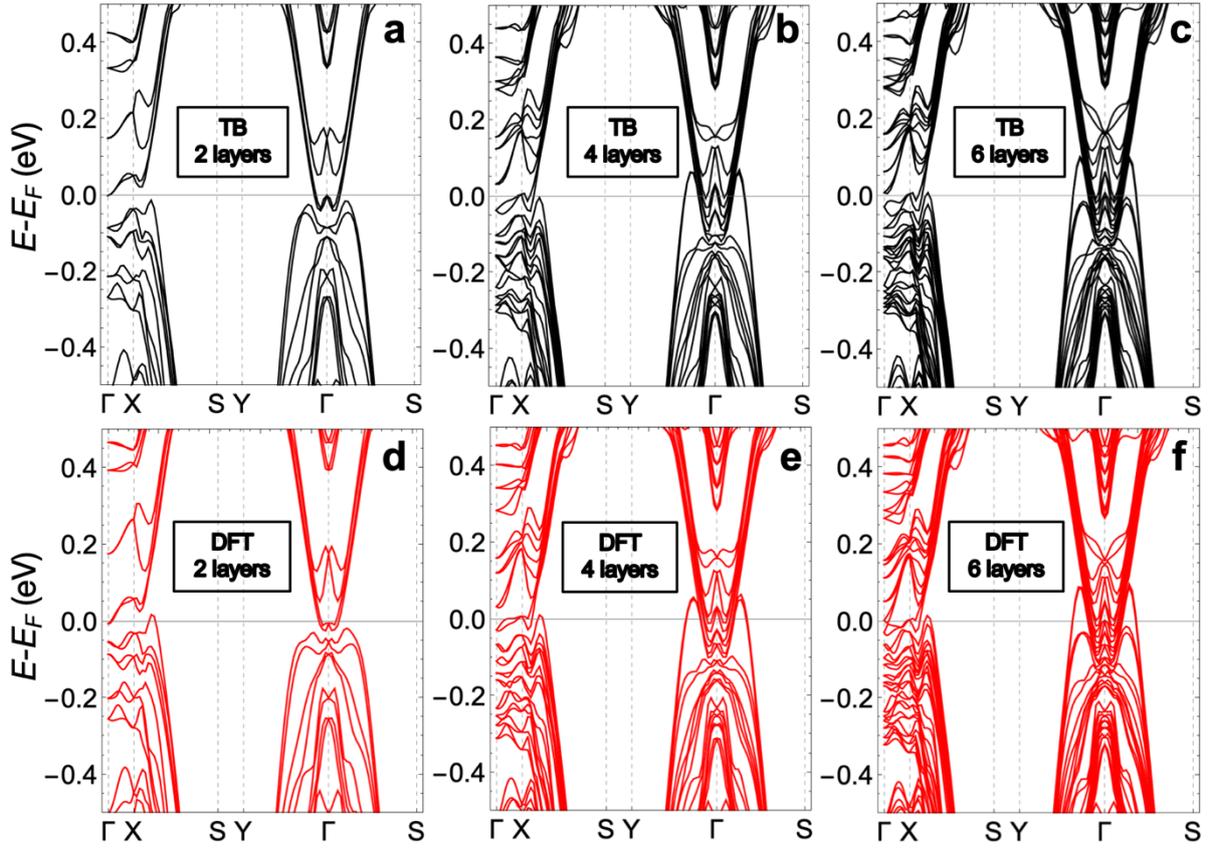

**Figure S9.** Comparison of slab band structures of supercell tight-binding (TB) and DFT calculations. The panel (a), (b), and (c) are TB band structures of two, four, and six-layer slabs calculated from supercell TB Hamiltonian, respectively, constructed from the bulk TB parameters. The panel (d), (e), and (f) are DFT band structures of two, four, and six-layer slabs, respectively, calculated with vacuum-slab geometry.

**Comparison of the band structures, BCD, and momentum-resolved BC between the GGA and LDA schemes**

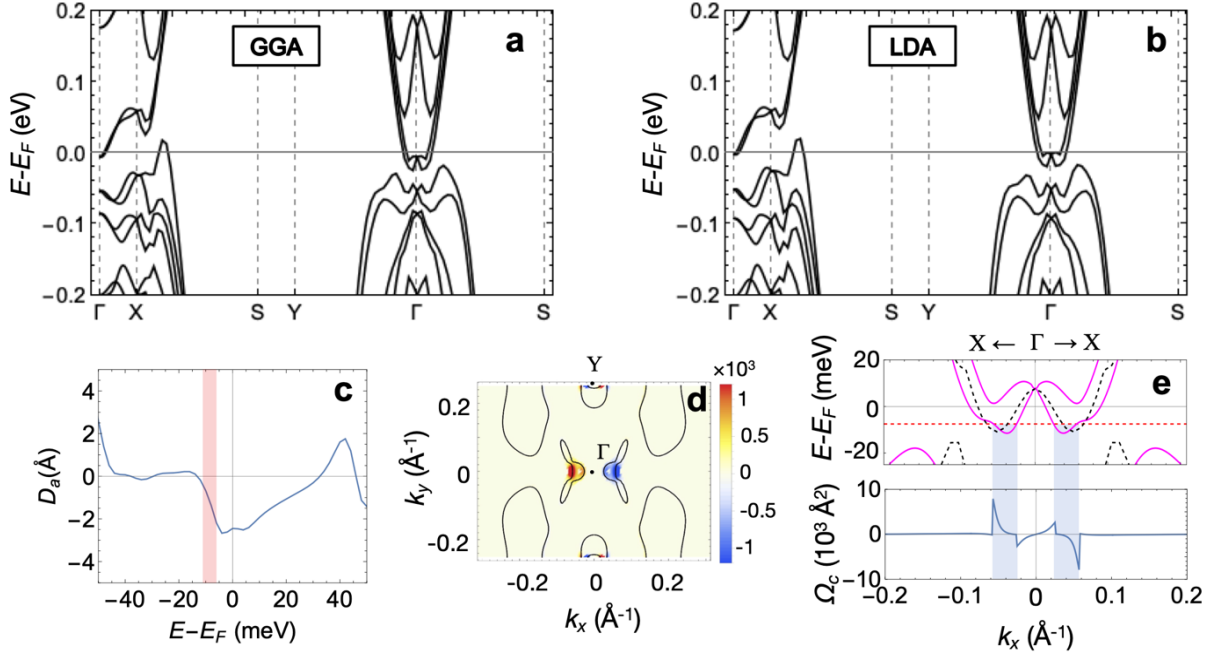

**Figure S10.** Comparison of BCD and momentum-resolved BC of the bilayer slab between LDA and GGA exchange-correlation functionals. **a-b.** Band structures along high-symmetry line calculated with GGA (panel a) and LDA (panel b) exchange-correlation functionals. The results in panels **c-e** are calculated with LDA. **c.** The BCD ($D_a$) contributing to the in-plane NLHE for bilayer NbIrTe$_4$, plotted as a function of the chemical potential. **d.** Momentum-resolved BC ($\Omega_c(\mathbf{k})$ in Å$^2$) of bilayer NbIrTe$_4$ integrated from −11 to −6 meV corresponding to the red shaded area in panel c. **e.** Band dispersion around the $k$-points contributing large Berry curvatures of bilayer NbIrTe$_4$. The top panel shows bands near the Fermi energy expressed as solid magenta lines along the $X$–$\Gamma$–$X$ cut. The black dashed lines are the bands calculated without spin-orbit coupling. The bottom panel is BC ($\Omega_c(\mathbf{k})$) integrated up to −7.5 meV relative to $E_F$ (red dashed line).

## Determination of the surface termination and doping-dependent surface band structures

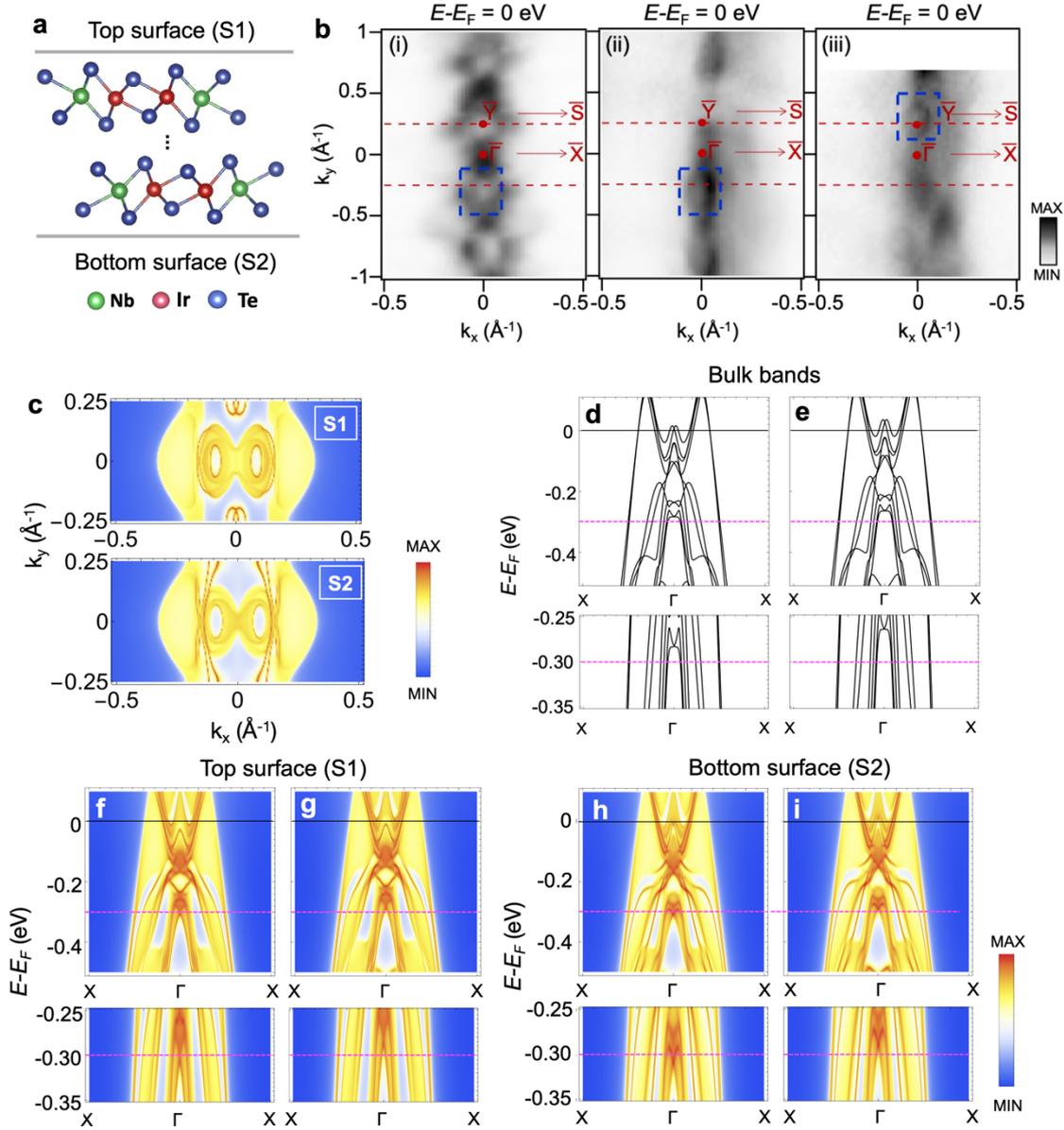

**Figure S11. Sample-dependent ARPES and surface band structures. a.** Schematics of the thin-film geometry showing top and bottom surfaces, denoted as S1 and S2, respectively. **b.** The Fermi surfaces obtained for three independent experiments (i)-(iii) using different samples, measured at 20 K, 20 K, and 10 K, respectively. The blue-dashed boxes show double hole pockets near the Y point that exist only on the S1 terminated surface. **c.** Fermi surfaces of surface band structures calculated for S1 (top panel) and S2 termination (bottom panel). Bulk band structures along Γ-X high symmetry line for undoped (**d**) and 0.025 h/f.u. doped (**e**) cases. The solid black and dashed magenta lines denote the Fermi energy and the binding energy of -0.3 eV. **f-i.** Surface band structures along Γ-X high symmetry line: (**f**) undoped with S1 termination, (**g**) 0.025 h/f.u. doped with S1 termination, (**h**) undoped with S2 termination, (**i**) 0.025 h/f.u. doped with S2 termination. The bottom panels of (**d-i**) show band structures around the -0.3 eV.

We have examined the surface termination for the ARPES measurements based on the surface-dependent features of the Fermi surface. There are two different top and bottom surface terminations in NbIrTe$_4$ as shown in Fig. S11a denoted by S1 and S2, respectively, following the definition of Ref. 29. We measure Fermi surfaces for three different samples as shown in Fig. S11b and find that all the samples show elliptical hole pockets at the Y point. To identify surface-dependent features in the ARPES data, we perform surface band structure calculations (panel **c**) that can be considered simulated ARPES measurements and find that only the S1 terminate surface has hole pockets at Y point, consistent with the Fermi surface in Ref 29. Thus, the comparison between measured and simulated ARPES spectra suggests that the surface of our samples shows the characteristics of the S1 termination.

In addition to identifying the surface termination, we check whether there are surface-dependent changes in the band structures induced by the chemical potential shift using theoretical calculations. The panels (**d-i**) in Fig. S11 present the bulk and surface band structures for undoped and hole-doped NbIrTe$_4$, in which the hole-doping effect is included by reducing the number of electrons (0.025 h/f.u.) with a compensating uniform background charge. The amount of doping is chosen such that the chemical potential shift of the bulk band structures is about 20 meV, close to the estimated value from ARPES (panels **d** and **e**).

We find that the surface band structures along the Γ-X high-symmetry line for both S1 and S2 terminations exhibit uniform upward shifts with amounts almost identical to the bulk bands. Thus, the temperature-dependent shift around -0.3 eV that we identify with ARPES spectra can be interpreted as the approximately same chemical potential shift, regardless of the surface termination.

**Doping-dependent electronic structures for the slab geometry**

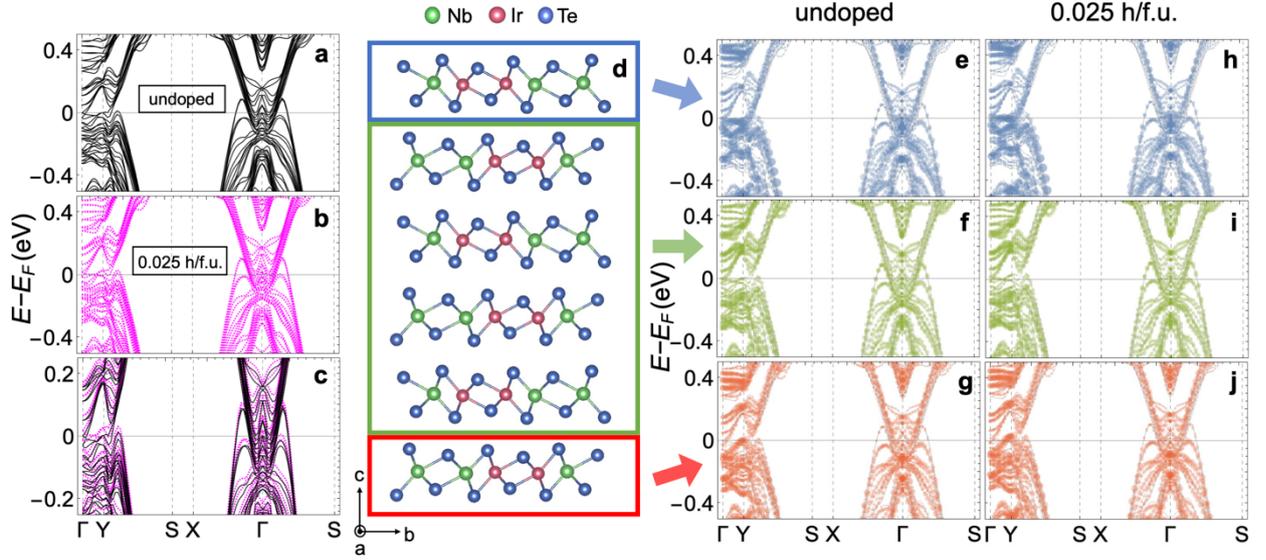

**Figure S12**. **Electronic structures of six-layer slab geometry for undoped and 0.025 h/f.u. doped NbIrTe$_4$.** Slab band structures of (**a**) undoped and (**b**) doped NbIrTe$_4$. **c.** Comparison between the undoped (black lines) and doped (magenta lines) band dispersions around the Fermi energy. **d.** Atomic configuration of the six-layer slab. The blue, green, and red boxed regions represent the top surface layer, middle layers, and bottom surface layer, respectively. **e-g.** Band structures of the undoped slab projected on the atoms in the top surface layer (panel **e**), middle layers (panel **f**), and bottom surface layer (panel **g**). **h-j.** Band structures of the doped slab projected on the atoms in the top surface layer (panel **h**), middle layers (panel **i**), and bottom surface layer (panel **j**).

     We calculate the electronic structures of the six-layer slab with undoped and doped (0.025 h/f.u.) cases. Figs. S12 (**a**) and (**b**) present the band structures of the undoped and doped six-layer slab geometry, respectively. We find that the effect of doping is mainly the rigid band shift of about 20 meV as shown in the comparison around the Fermi energy (panel (**c**)), which is consistent with our bulk calculations.

    In addition, we evaluate the projected bands on the atoms in the top and bottom layers and their doping dependence, since surface regions lacking the glide mirror symmetry ($\{M_b|(1/2,0,1/2)\}$) of the bulk structure mainly contribute to BCD$^2$. Fig. S12 (**d**) shows the atomic configuration of the six-layer slab, where we divide the atomic structure into the top surface layer, middle bulk-like layers, and bottom surface layer, shown in blue, green, and red boxes, respectively. Figs. S12 (**e**), (**f**), and (**g**) are bands projected on the atoms belonging to the top, middle, and bottom layers, respectively, and similarly, Figs S12 (**h-j**) are corresponding atom-projected bands for the doped case. We find no significant differences between the

projected bands among the three regions, consistent with the relatively weak inter-layer van der Waals coupling. More importantly, we find no noticeable changes in the projected bands with respect to doping, suggesting that the main effect of doping is a rigid shift in the band structures.